\documentclass[conference]{IEEEtran}
\IEEEoverridecommandlockouts
\usepackage{cite}
\usepackage{amsmath,amssymb,amsfonts}
\usepackage{algorithmic}
\usepackage{graphicx}
\usepackage{textcomp}
\usepackage{xcolor}

\usepackage{float}
\usepackage{acronym}
\usepackage{eurosym}
\usepackage{tikz}

\usepackage{siunitx} 
\DeclareSIUnit{\atm}{atm}
\DeclareSIUnit{\kWh}{kWh}
\DeclareSIUnit{\Ah}{Ah}
\DeclareSIUnit{\eur}{\mbox{\text{\euro}}}
\DeclareSIUnit{\kW}{kW}
\DeclareSIUnit{\MW}{MW}
\DeclareSIUnit{\kwm}{kW/m^2}
\DeclareSIUnit{\m}{m}
\def\BibTeX{{\rm B\kern-.05em{\sc i\kern-.025em b}\kern-.08em
    T\kern-.1667em\lower.7ex\hbox{E}\kern-.125emX}}

\acrodef{bess}[BESS]{Battery Energy Storage System}
\acrodef{ec}[EC]{European Commission}
\acrodef{ess}[ESS]{Energy Storage System}
\acrodef{rec}[REC]{Renewable Energy Community}
\acrodefplural{rec}[RECs]{Renewable Energy Communities}
\acrodef{res}[RES]{Renewable Energy Source}
\acrodef{soc}[SoC]{State of Charge}
\acrodef{pv}[PV]{Photovoltaic}
\acrodef{der}[DER]{Distributed Energy Resource}
\acrodef{tso}[TSO]{Transmission System Operator}
\acrodef{msd}[ACS]{Ancillary Service Market}
\acrodef{mise}[MISE]{Italian Ministry of Economic Development}
\acrodef{dmc}[DMC]{Discrete Markov Chain}
\acrodef{cf}[CF]{Community Facility}
\newcommand\copyrighttext{%
  \footnotesize
  \centering\copyright~2023 IEEE. Personal use of this material is permitted. Permission from IEEE must be obtained for all other uses, in any current or future media, including reprinting/republishing this material for advertising or promotional purposes, creating new collective works, for resale or redistribution to servers or lists, or reuse of any copyrighted component of this work in other works. \\ Submitted at the IEEE PES General Meeting 2023.}
\newcommand\copyrightnotice{%
\begin{tikzpicture}[remember picture,overlay]
\node[anchor=south,yshift=0pt] at (current page.south) {\setlength{\fboxrule}{0pt}\fbox{\parbox{\dimexpr\textwidth-\fboxsep-\fboxrule\relax}{\copyrighttext}}};
\end{tikzpicture}%
}

\begin{document}

\title{Day-Ahead Programming of Energy Communities Participating in Pay-as-Bid Service Markets\thanks{Funded by the European Union - Next Generation EU - NRRP M4.C2 - Investment 1.5 Establishing and strengthening of Innovation Ecosystems for sustainability (Project n. ECS00000024, Rome Technopole).}}

\author{%
  \IEEEauthorblockN{F. Conte}
  \IEEEauthorblockA{Univesity Campus Bio-Medico of Rome\\
    Faculty of Engineering\\
    Via Alvaro del Portillo, 21\\
    I-00128, Roma, Italy\\
    f.conte@unicampus.it}
     \and
 \IEEEauthorblockN{S. Massucco, G. Natrella, M. Saviozzi, F. Silvestro}
 \IEEEauthorblockA{University of Genoa\\
    DITEN\\
    Via all'Opera Pia 11 A\\
    I-16145, Genova, Italy\\
    stefano.massucco@unige.it}}

\IEEEaftertitletext{\copyrightnotice\vspace{1.1\baselineskip}}
\maketitle
\begin{abstract}
This paper proposes an optimal strategy for a Renewable Energy Community participating in the Italian pay-as-bid ancillary service market. The community is composed by a group of residential customers sharing a common facility equipped with a PV power plant and a battery energy storage system. This battery is employed to maximize the community cash flow obtained by the participation in the service market. A scenario-based optimization problem is defined to size the bids to be submitted to the market and to define the corresponding optimal battery energy exchange profile for the day ahead. The proposed optimization scheme is able to take in to account the probability of acceptance of the service market bids and the uncertainties on the forecasts of PV generation and energy demands. Results prove the effectiveness of the approach and the economic advantages on participating in the service market. 
\end{abstract}



\section{Introduction}\label{sec:Introduction}
Among the solutions proposed to integrate \acp{res} in the power system, \acp{rec}, defined by the EU directive \cite{REDII}, have emerged as a dynamic and multifaceted entity composed by residential customers, small-medium enterprises and/or local authorities, engaging in various activities centered on the ownership and management of \acp{res}. The primary aim of \acp{rec} is to generate, store, and sell renewable energy by facilitating the exchange of this energy among its members.  


In the Italian law transportation \cite{it-trans} of the EU directive, \acp{rec}' members are encouraged to share renewable energy through the payment of an incentive, but they also have the opportunity to be a player in the \ac{msd}, which follows a pay-as-bid mechanism. 
In literature, different techniques are proposed to manage \ac{res} generation and storage systems to optimize their functioning in the context of \acp{rec} compliant with the Italian law such as \cite{conteEAI,bianchiAE,barone2020smart}. Few works optimize \acp{rec} for multi-market participation. A bidding-based peer-to-peer energy transaction optimization model is presented in \cite{park}. The proposed model considers \ac{rec} members willingness to pay for green energy and minimizes the community operating costs. A two-stage stochastic programming model for energy trading, evaluating the potential to supply ancillary services is presented in \cite{GARCIAMUNOZ}. However, in literature there is a lack of papers addressing the problem of \ac{rec} optimal participation in pay-as-bid service markets. 

In this paper we provide an optimal scenario-based day-ahead programming algorithm that optimizes the participation of \ac{rec} in the Italian \ac{msd}. As shown in Figure~ \ref{fig:schema}, the \ac{rec} is composed by a group of residential customers sharing a \ac{cf} equipped with a \ac{pv} power plant, a \ac{bess} and an internal load. The proposed algorithm allows the \ac{rec} to optimally size the bids to be submitted to the \ac{msd} and to define the \ac{bess} energy exchange for the day ahead, taking into account the uncertainties on the forecasts of \ac{pv} generation and energy demands, as well as the probability of acceptance of the \ac{msd} bids.  Simulation results that prove the effectiveness of the approach and the potential economic advantages for a \ac{rec} in participating in the \ac{msd}.



\section{System Architecture and Market Rules}\label{sec:SystemArchitecture}

\begin{figure}[t]
	\centering
    \includegraphics[width=0.9\columnwidth]{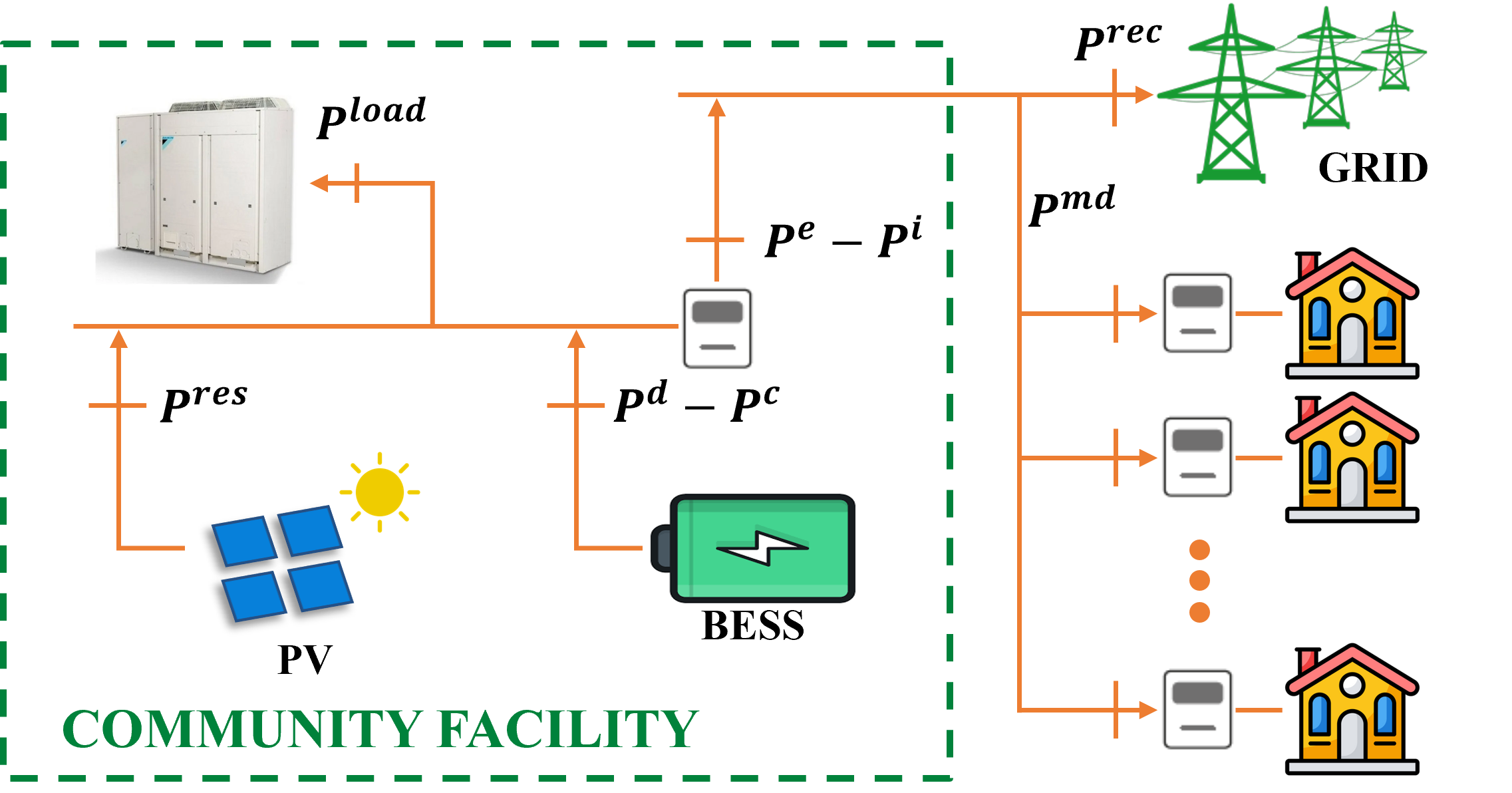}
    	\caption{REC architecture.}
	\label{fig:schema}
\end{figure}

The schematic architecture of the considered \ac{rec} is shown in Figure \ref{fig:schema}. 
Here, links do not necessarily compose an actual electrical network since, according to the Italian law \cite{it-trans}, the energy exchange within a \ac{rec} can be virtual, with the unique condition that \acp{res} and consumers are connected under the same HV/MV substation. 

\subsection{Energy Markets Rules}\label{ssec:market_rules}
During any hour $k=1,2,\ldots,24$, the \ac{cf} can export energy $P^e_k$ or import energy $P^i_k$ with the utility grid in order to meet its internal demand $P^{load}_k$ and share energy with the \ac{rec} members. According to the Italian law \cite{it-trans}, the \ac{cf} does not have to fully meet the aggregated members demand $P^{md}_k$, as every member keep their independence and contract with their energy provider. As a virtual entity, the \ac{rec} energy exchange with the utility grid is 
\begin{equation} \label{eq:rec_exchange}
 P^{rec}_k = P^e_k-P^i_k-P^{md}_k   
\end{equation}
with positive values meaning export.

The \ac{rec} pays $c^i_k$ for the energy imported by the \ac{cf} from the utility grid and is paid $c^e_k$ for the energy exported by the \ac{cf} to the utility grid. According to the Italian law \cite{it-trans}, the \ac{rec} gets an incentive $\gamma^{sh}$ proportional to the \textit{shared energy} $P^{sh}$, defined as the minimum between the aggregated members energy demand and the exported renewable energy, \textit{i.e.}
\begin{equation}
    P^{sh}_k = \min\{P^e_k,P^{md}_k\} \label{eq:shared}.
\end{equation} 

In this paper we consider the participation of the \ac{rec} in the Italin ac{msd} where
the \ac{tso} acts as the central counterpart. When necessary, the \ac{tso} requires participants to modify their baseline, declared the day before, by generate or consume more/less power. In the \ac{msd} market sessions, participants declare the action that they are willing to realize and submit their prices for that. The \ac{tso} accept bids according to a pay-as-bid mechanism. 

More specifically, \ac{rec}'s bid in the \ac{msd}, when accepted, is acknowledged by an economic reward \textit{i.e.}, the \ac{rec} is willing to modify its baseline and it is paid to generate more (or consume less) energy (sell bid) or, on the contrary, pays to generate less (or consume more) energy (purchase bid). Bids are composed by a couple price-energy variation, $(c^+_k,P^+_k)$ for sell bids and $(c^-_k,P^-_k)$ for purchase bids. Service variations $P^+_k$ and $P^-_k$ are defied with respect a baseline $\hat{P}_k^{rec}$ that the \ac{rec} must declare together with the bids. 

We assume that the \ac{tso} is price-taker with respect to the \ac{rec}. This means that a sell bid is fully accepted when the offered price $c^+_k$ is lower than the maximum accepted sell price in the \ac{msd} $\overline{c}^{+}_k$, while a purchase bid is fully accepted when the offered price $c^-$ is higher than the minimum accepted purchase price in the \ac{msd} $\underline{c}^-_k$. Clearly, $\overline{c}^{+}_k$ is usually higher than $c^e$, as well as, $\underline{c}^-_k$ is usually lower than $c^i_k$.

Therefore, if at any hour $k$ a sell bid is accepted the \ac{rec} must exchange with the grid the energy $\hat{P}_k^{rec}+P^+_k$. In the case that $P^{rec}_k<\hat{P}_k^{rec}+P^+$, the error $E^+_k = \hat{P}_k^{rec}+P^+_k - P^{rec}_k$ is paid by the \ac{rec} to the \ac{tso} with a penalty $p^+$ higher than $\overline{c}^{+}_k$. Similarly, if a purchase bid is accepted the \ac{rec} energy exchange with the grid should be $\hat{P}_k^{rec}-P^-_k$ and in the case that $P^{rec}_k>\hat{P}_k^{rec}-P^-_k$, the service error $E^-_k = P^{rec}_k-(\hat{P}_k^{rec}-P^-_k)$ is paid by the \ac{tso} to the \ac{rec} with a price $p^-$ lower than $\underline{c}^-_k$.

\subsection{Objective}\label{ssec:problem_formulation}
The objective of this paper is to program, for all hours $k=1,2,\ldots,24$ of the day- ahead, the \ac{rec} baseline $\hat{P}^{rec}_k$ and the sell and buy bids to be submitted to the \ac{msd}, $(c^+_k,P^+_k)$ and $(c^-_k,P^-_k)$ assuming to be able to control the community \ac{bess} power exchange and being provided with: 1) initial \ac{soc} of the \ac{bess} $S_0$; 2) forecasts the market prices $c^e_k$, $c^i_k$, $\overline{c}^+_k$ and $\underline{c}^-_k$; 3) forecasts of renewable generation and demands, $P^{res}_k$, $P^{load}_l$ and $P^{md}_k$. 

The program objective is to maximize the day cash flow of the community. When the day-ahead optimization is processed, this cash flow is a random variable since we are provided with forecasts of prices, generation and demands. Therefore, we actually realize stochastic optimization where the expected value of the day cash flow is maximized.

\section{Forecasting Methods}
\subsection{Energy Prices Forecast} \label{ssec:forecast_prices}
We first suppose that, when the day-ahead programming is computed, import and export prices $c^i_k$ and $c^e_k$ with $k=1,2,\ldots,24$ are known. Differently, \ac{msd} prices cannot be known a-priori. Therefore, a scenario based forecast approach is adopted. Hystorical data of the maximum accepted sell prices $\overline{c}^{+}_k$ and minimum accepted purchase prices $\underline{c}^{-}_k$ (available on the Italian electricity market operator web repository \cite{gme}) are collected by considering each couple of daily trajectories of $\overline{c}^{+}_k$ and $\underline{c}^{-}_k$ as one scenario. Then, using the scenario reduction algorithm in~\cite{Growe:2003} a reduced set of $n_m$ scenarios $\{\{\overline{c}^{+}_{k,s}\}_{k=1}^{24},\{\underline{c}^{-}_{k,s}\}_{k=1}^{24}\}_{s=1}^{n_m}$ is obtained, each associated to the probability $\pi^m_s$, with $s=1,\ldots,n_m$.

\subsection{Demands and \ac{res} Generation Forecasts} \label{ssec:forecasts}
In~\cite{PMAPS2020}, a methodology is proposed to model time series with a \ac{dmc} through an adaptive online algorithm with minimal computational efforts. The constructed \ac{dmc} can then be used to sample possible future scenarios given the current actual state of the \ac{dmc}. The algorithm for the case of three variables \textit{i.e.}, \ac{res} generation $P^{res}$ and energy demands $P^{md}$ and $P^{load}$ is detailed in~\cite{conteEAI}. The \ac{dmc} is used to generate 300 scenarios with a time length of 24 hours. Then, the scenario reduction algorithm in~\cite{Growe:2003} is used to reduce the number of scenarios to $n_r$, each one with its own associated probability. The results are the forecast scenarios $\{\{P^{res}_{k,l}\}_{k=1}^{24},\{P^{load}_{k,l}\}_{k=1}^{24},\{P^{md}_{k,l}\}_{k=1}^{24}\}_{l=1}^{n_r}$, associated to the probability $\pi^r_{l}$, with $l=1,\ldots,n_r$.

\section{Day-Ahead Programming}\label{sec:Algorithm}
As stated in Section~\ref{ssec:problem_formulation}, our objective is establish a day-ahead program for the \ac{cf} with the objective of maximizing the expected value of the community cash flow. Since random time-series are involved, we formulate a stochastic optimization problem. Our choice is to carry out a scenario based optimization \cite{scenario-based}, where random time-series distributions are approximated by a set of scenario trajectories, each associated to an occurrence probability. As detailed in Section~\ref{ssec:forecast_prices} and Section~\ref{ssec:forecasts} the forecasts of all random time-series involved in our problem are built as scenarios. We have $n_m$ \textit{prices scenarios}, indexed by $s=1,2,\ldots,n_m$ (Section~\ref{ssec:forecast_prices}), and $n_r$ \textit{energy scenarios} (for \ac{pv} generation and demands, Section~\ref{ssec:forecasts}), indexed by $l=1,2,\ldots,n_r$. In the following, if the value of a given variable $x_k$ depends on the realizations of the prices or the energy scenarios, we will indicate the corresponding realization of $x_k$ with $x_{k,s}$ or with $x_{k,l}$. In the case $x_k$ depends on both the scenarios' realizations, the corresponding realization of $x_k$ will be indicated with $x_{k,s,l}$. 

In order to define the day-ahead optimization problem we need to define a proper objective function and introduce a set of auxiliary variables and mixed-integer constraints modeling the \ac{msd} mechanism and the desired behaviour of the \ac{rec}.

First of all we introduce two binary variables $b_k^{+/-}$ equal to 1 if the \ac{rec} decides to submit a sell or a purchase bid at time $k$, respectively. Therefore:
\begin{align}
    0\leq P^+_k \leq P^e_{max}&b^+_k, \quad \ 0\leq P^-_k \leq P^i_{max}b^-_k \label{eq:bplus} \\
    &b^+_k + b^-_k \leq 1 \label{eq:bplusminus}
\end{align}
where $P_{max}^{e}$ and $P_{max}^{i}$ are the maximum powers that \ac{cf} can export to and import from the grid, respectively. Constraint \eqref{eq:bplusminus} is introduced to impose that at a given time $k$ there be only one bid, either to sell or to purchase.

Let us now define two random variables $\Gamma_{k,s}^+$ and $\Gamma_{k,s}^-$ as the upward and downward variations that the \ac{rec} must realize if the sell and purchase bids at time $k$ are accepted in the prices scenario $s$. Thus, according to the rules introduced in Section~\ref{ssec:market_rules}, for all $s=1,2,\ldots,n_m$,
\begin{align}
    & \Gamma^+_{k,s} = \delta^+_{k,s} {P}^{+}_{k} \label{eq:gamma1}\\
    & \delta^+_{k,s} = 1  \Leftrightarrow c^{+}_{k} \leq \overline{c}^+_{k,s} \label{eq:gamma2} \\
    & \Gamma^-_{k,s} = \delta^-_{k,s} {P}^{-}_{k}\label{eq:gamma3} \\
    & \delta^-_{k,s} = 1  \Leftrightarrow c^{-}_{k} \geq \underline{c}^{-}_{k,s} \label{eq:gamma4}
\end{align}
where $\delta^{+/-}_{k,s}$ are binary variables equal to one if, in the prices scenario $s$, sell or purchase bids are accepted.

Let us consider now the energy exchange of the \ac{cf} and of the overall \ac{rec}. If a sell (or purchase) bid is accepted, the upward (or downward) energy variation should be realized by the \ac{cf} by using the energy reserve kept by the \ac{bess}. Moreover, we assume to use the \ac{bess} also to compensate the forecasting errors on \ac{pv} generation and demands. This means that the energy exchanges of \ac{rec} and \ac{cf} will depend both on the prices and energy scenarios, meaning that $P^{rec}_k$, $P^i_k$ and $P^e_k$ become $P^{rec}_{k,s,l}$, $P^i_{k,s,l}$ and $P^e_{k,s,l}$. Under these assumptions, the \ac{cf} energy balance is modeled by: 
\begin{align}
& P^e_{k,s,l} -  P^i_{k,s,l} = P^{res}_{k,l}-P^{load}_{k,l} + P^{d}_{k,s,l} -P^{c}_{k,s,l} \label{eq:cf1} \\
& 0\leq {P}^e_{k,s,l} \leq {\delta}^e_{k,s,l}P^e_{max}\\
& 0\leq {P}^i_{k,s,l} \leq (1-{\delta}^e_{k,s,l})P^i_{max} \label{eq:cf2a}
\end{align}
where $\delta^e_{k,s,l}$ are binary variable equal to 1 if the \ac{cf} is exporting energy. Moreover, the \ac{rec} energy exchange \eqref{eq:rec_exchange} is represented as follows:
\begin{align}
& P^{rec}_{k,s,l} = P^e_{k,s,l}-P^i_{k,s,l}-P^{md}_{k,l} \label{eq:rec1}\\
&\begin{aligned} 
&P^{rec}_{k,s,l} = \hat{P}^{rec}_{k}+w^+_{k,s,l}-w^+_{k,s,l}\\
&\qquad \qquad +\Gamma^+_{k,s} - E^+_{k,s,l} - (\Gamma^-_{k,s} - E^-_{k,s,l})
\end{aligned}\label{eq:rec2}\\
& 0 \leq E^+_{k,s,l}\leq \Gamma^+_{k,s,l},\quad 0\leq E^-_{k,s,l}\leq \Gamma^-_{k,s,l} \label{eq:rec4}
\end{align}
In \eqref{eq:rec2} we impose the desired energy balance of the \ac{rec} for all possible scenarios, \textit{i.e.} variation from the baseline $\hat{P}^{rec}_k$ when one sell or purchase bid is submitted and accepted. $E^{+/-}_{k,s,l}$ can be considered as constraint relaxing variables, recalling that, if different from zero, the \ac{rec} will pay penalties. While, $w_{k,s,l}^{+/-}$ are relaxing variable allowing the baseline to be modified according to the random variations of generation and demands in the scenarios where no bid is accepted.

We need now to impose that when bids are accepted, the service variations are realized by the \ac{cf} using the \ac{bess}. Therefore, the following constraints must hold true: 
\begin{align}
&\begin{aligned} 
&P^{e}_{k,s,l}-P^{i}_{k,s,l} = \hat{P}^{i}_{k,s,l} -\hat{P}^{e}_{k,s,l}\\
&\qquad \qquad \qquad \quad +\Gamma^+_{k,s} - E^+_{k,s,l} - (\Gamma^-_{k,s} - E^-_{k,s,l})
\end{aligned}\label{eq:cf3}\\
& 0\leq \hat{P}^e_{k,s,l} \leq \hat{\delta}^e_{k,s,l}P^e_{max}\label{eq:cf4}\\
& 0\leq \hat{P}^i_{k,s,l} \leq (1-\hat{\delta}^e_{k,s,l})P^i_{max} \label{eq:cf5}\\
& P^{d}_{k,s,l} -P^{c}_{k,s,l} = \hat{P}^{b}_{k} +  \Gamma^+_{k,s,l} - \Gamma^-_{k,s,l} + v^+_{k,s,l} -  v^-_{k,s,l}\label{eq:cf6} \\
& 0\leq {P}^c_{k,s,l} \leq P^b_{nom} {\delta}^c_{k,s,l} \\  
&0\leq {P}^d_{k,s,l} \leq P^b_{nom} (1-{\delta}^c_{k,s,l})  \label{eq:cf7}
\end{align}
where $\hat{P}^{i}_{k,s,l} -\hat{P}^{e}_{k,s,l}$ is the \ac{cf} baseline called to realize the \ac{rec} baseline $\hat{P}^{rec}_{k}+w^+_{k,s,l}-w^-_{k,s,l}$ in \eqref{eq:rec2} when no bid is accepted. Constraint \eqref{eq:cf6} imposes that the service variations are effectively provided by the \ac{bess}. Here, $\hat{P}^{b}_{k}$ is the \ac{bess} baseline, followed by the \ac{bess} when no bid is accepted, whereas $v_{k,s,l}^{+/-}$ are relaxing variables which allow the \ac{bess} to be provided with a reserve to compensate the energy forecasting errors and reduce, in this way, errors $E^{+/-}_{k,s,l}$. 

More specifically, the use of the relaxing variables $w_{k,s,l}^{+/-}$ and $v_{k,s,l}^{+/-}$ is regulated by the following constraints:
\begin{align}
& \delta^+_{k,s}+b^+_k=2  \Leftrightarrow  u^+_{k,s} = 1\label{eq:con15}\\
& \delta^-_{k,s}+b^-_k=2  \Leftrightarrow  u^-_{k,s} = 1\label{eq:con16}\\
&  0 \leq w^+_{k,s,l} \leq \varepsilon_{max}(1-u^-_{k,s}) \label{eq:con17}\\
&  0 \leq w^-_{k,s,l} \leq \varepsilon_{max}(1-u^+_{k,s}) \label{eq:con18}\\
&  0 \leq v^+_{k,s,l} \leq \varepsilon_{max}u^+_{k,s} \label{eq:con19}\\
&  0 \leq v^-_{k,s,l} \leq \varepsilon_{max}u^-_{k,s} \label{eq:con20}
\end{align}
where $\varepsilon_{max}$ is the expected maximal range of variation of the balance $P^{res}-P^{load}-P^{md}$. In particular, thanks to \eqref{eq:con15}--\eqref{eq:con16}, binary variables $u_{k,s}^+$ and $u^-_{k,s}$ identify if in scenario $s$ sell or purchase bids have been submitted and accepted. Then, constraints \eqref{eq:con17}--\eqref{eq:con20} suitably deactivate the use of the relaxing variables according to the following rules: in the case a sell bid is accepted, if the balance $P^{res}-P^{load}-P^{md}$ is lower then expected, the \ac{bess} is called to augment its energy export (or reduce its energy import) to reduce the service error $E^+$: thus, $v^+_{k,s,l}>0$ by \eqref{eq:con19} and $w^-_{k,s,l}=0$ by \eqref{eq:con18}. 
Specularly, in the case a purchase bid is accepted, if the balance $P^{res}-P^{load}-P^{md}$ is higher then expected, the \ac{bess} is called to reduce its energy export (or augment its energy import) to reduce the service error $E^-$: thus, $v^-_{k,s,l}>0$ by \eqref{eq:con20} and $w^+_{k,s,l}=0$ by \eqref{eq:con17}. 
According to \eqref{eq:cf6}, the \ac{bess} energy exchange will depend on the prices and energy scenarios. Therefore, we need to set scenario depending constraints on the \ac{bess} \ac{soc}:
\begin{align}
    & 0\leq S_0 + \sum_{j=1}^{k} \frac{1}{E_b}\left(\eta_c P^c_{j,s,l}-\frac{1}{\eta_d} P^d_{j,s,l}\right)\leq 1 \label{eq:con_soc}\\
    &S^{24}_{min}\leq S_0 + \sum_{j=1}^{24} \frac{1}{E_b}\left(\eta_c P^c_{k,s,l}-\frac{1}{\eta_d} P^d_{k,s,l}\right)\leq S^{24}_{max}\label{eq:con_soc24} \\
    &0 \leq P^c_{k,s,l} \leq P^b_{nom} \delta^c_{k,s,l}, 0\leq P^d_{k,s,l} \leq P^b_{nom} (1-\delta^c_{k,s,l}) \\
    &P^c_{k,s,l} \leq P^{res}_{k,l} \label{eq:green}
\end{align}
where: $E_b$ is the battery capacity; $\eta_{c}$ and $\eta_{d}$ are the charge and discharge efficiencies, respectively; $P^{c}_{k,s,l}$ and $P^{d}_{k,s,l}$ are charging (import) and discharging (export) energies, respectively; $\delta_{k,s,l}^c$ are binary variable equal to 1 when battery charging is active; $P^b_{nom}$ is the nominal battery power; and $[S^{24}_{min}, S^{24}_{max}]$ is the interval within which the \ac{soc} must terminate at the end of the day. 
Constraint \eqref{eq:green} ensures that the \ac{bess} is charged only with renewable energy an thus exported energy is renewable as well.

Before introducing the objective function and finally obtain our day-ahead optimization problem, we implement the definition of the shared energy \eqref{eq:shared} by the following constraints:
\begin{equation}
  P^{sh}_{k,s,l}\leq P_{k,l}^{rec}, \quad P^{sh}_{k,s,l} \leq P^{e}_{k,s,l}\label{eq:con_shared}  
\end{equation}
where $P^{sh}_{k,s,l}$ is the shared energy obtained with the occurrence of the price scenario $s$ and the energy scenario $l$. Constraints \eqref{eq:con_shared} are sufficient to implement definition \eqref{eq:shared} since they are linear inequalities and $P^{sh}_{k,s,l}$ will be maximized according to the objective function \eqref{eq:obj_func}, introduced in the following.

We finally introduce the objective function to be maximized, defined as the expected value of the community cash flow:
\begin{equation}
\begin{aligned}\label{eq:obj_func}
    J=& \sum_{k=1}^{24} \sum_{s=1}^{n_m}  \sum_{l=1}^{n_r} \pi^m_s \pi^r_l \left(\hat{P}^e_{k,s,l}c^e_k - \hat{P}^i_{k,s,l}c^i_k +P^{sh}_{k,s,l}\gamma^{sh}  \right. \\
    &\qquad \quad \left. + \Gamma^+_{k,s} c^{+}_{k} - E^+_{k,s,l} p^+ - \Gamma^-_{k,s} c^{-}_{k} + E^-_{k,s,l} p^- \right)
\end{aligned} 
\end{equation}

Because of terms $\Gamma^+_{k,s} c^{+}_{k}$ and $\Gamma^+_{k,s} c^{+}_{k}$, \eqref{eq:obj_func} results to be a non convex quadratic function. This issue can however overcome by imposing that the bid prices $c^{+/-}_{k}$ are equal to one of the predicted prices $\overline{c}^+_{k,s}$ and $\underline{c}^+_{k,s}$ or to zero, \textit{i.e.} by substituting $c^{+/-}_{k}$ with 
\begin{align}
    & c^{+}_{k} = \sum_{j=1}^{n_m} d^{+}_{k,j} \overline{c}^{+}_{k,j}, \quad \sum_{j=1}^{n_m} d^{+}_{k,j} = b^+_{k} \label{eq:con_c+} \\
    & c^{-}_{k} = \sum_{j=1}^{n_m} d^{-}_{k,j} \underline{c}^{-}_{k,j}, \quad \sum_{j=1}^{n_m} d^{-}_{k,j} = b^-_{k} \label{eq:con_c-}
\end{align}
where $d^{+/-}_{k,j}$ are binary variables. In this way, objective function \eqref{eq:obj_func} becomes linear mixed-integer.

The optimization problem to be solved to obtain the day-ahead programming of the \ac{rec} is finally formulated as:
\begin{equation} \label{eq:opt_prob}
    \max J \ \ \mbox{s.t. \eqref{eq:bplus}-\eqref{eq:con_shared}}
\end{equation}
Notice also that relations \eqref{eq:gamma2}, \eqref{eq:gamma4}, \eqref{eq:con15} and \eqref{eq:con16} can be rewritten as a set of linear inequalities following, for example, the rules introduced in \cite{bemporad}. Therefore, the optimization problem \eqref{eq:opt_prob} results to be linear mixed-integer, ready to be solved with a proper solver. In the simulations presented in the next section, we used Gurobi \cite{gurobi}.

\section{Results}\label{sec:Results}
In order to test the performance of the proposed day-ahead programming scheme, we simulated a \ac{rec} with the structure in Figure~\ref{fig:schema} with 15 residential customers having an aggregated nominal demand of 37.076~kW, a 50~kWp \ac{pv} plant, an internal load with nominal power 10.037~kW and a 120~kW~/~250~kWh \ac{bess}. The maximal import/export of the \ac{cf} is $P^i_{max}=P^e_{min}=200$~kW. \ac{bess} efficiencies are set to $\eta^c=\eta^d=0.95$, and $S^{24}_{min}$ and $S^{24}_{max}$ to 0.3 and 0.7, respectively. Data from \cite{sergioramos} have been used for \ac{pv} generation and residential customers and internal load energy demands.

We simulated one week, considering the energy prices of the first week of July 2022, downloaded from the Italian electricity market operator web repository \cite{gme}. Prices scenarios have been computed as detailed in Section~\ref{ssec:forecast_prices}, using data from a month preceding the simulated week as starting data-set, then reduced to $n_m = 10$ scenarios. The incentive for shared energy is fixed to $\gamma_{sh}=0.119$~\euro/kWh, according to the Italian law \cite{it-trans}. For energy scenarios, the method mentioned in Section~\ref{ssec:forecasts} was adopted. The \ac{dmc} was trained with historical data in \cite{sergioramos}. The final number of scenarios is $n_r=10$. 

\begin{figure}[t]
	\centering
    \includegraphics[width=0.9\columnwidth]{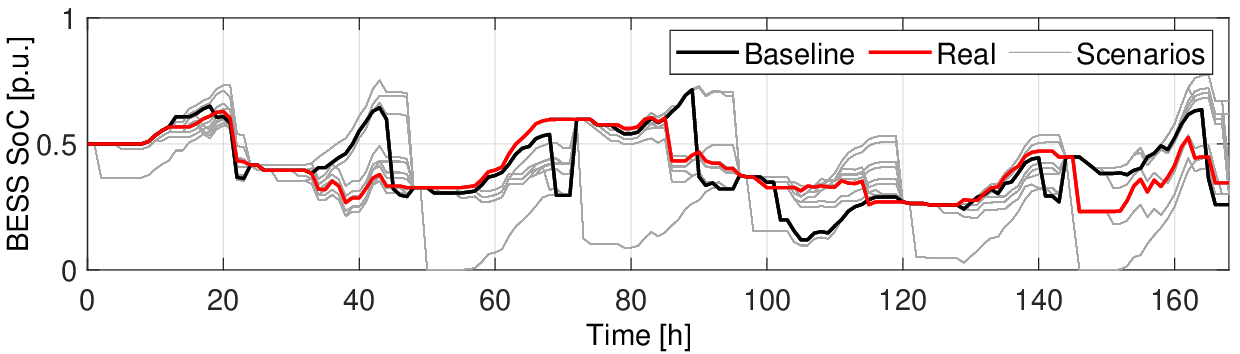}\vspace{-5pt}
    	\caption{Results: \ac{bess} \ac{soc} profile.}  
	\label{fig:soc}
\end{figure}

Figure~\ref{fig:soc} illustrates the \ac{soc} of the \ac{bess}, during the week. Grey lines are the expected scenarios \ac{soc} profiles, defined according to \eqref{eq:con_soc}--\eqref{eq:con_soc24}, by which the optimization explores the flexibility potentially provided by the \ac{bess}. The consistency of the proposed algorithm is proved by the fact that, as expected, the real \ac{soc} obtained during operation (red line) is always in between the bounds defined by the scenarios profiles. 

\begin{figure}[t]
	\centering
    \includegraphics[width=0.9\columnwidth]{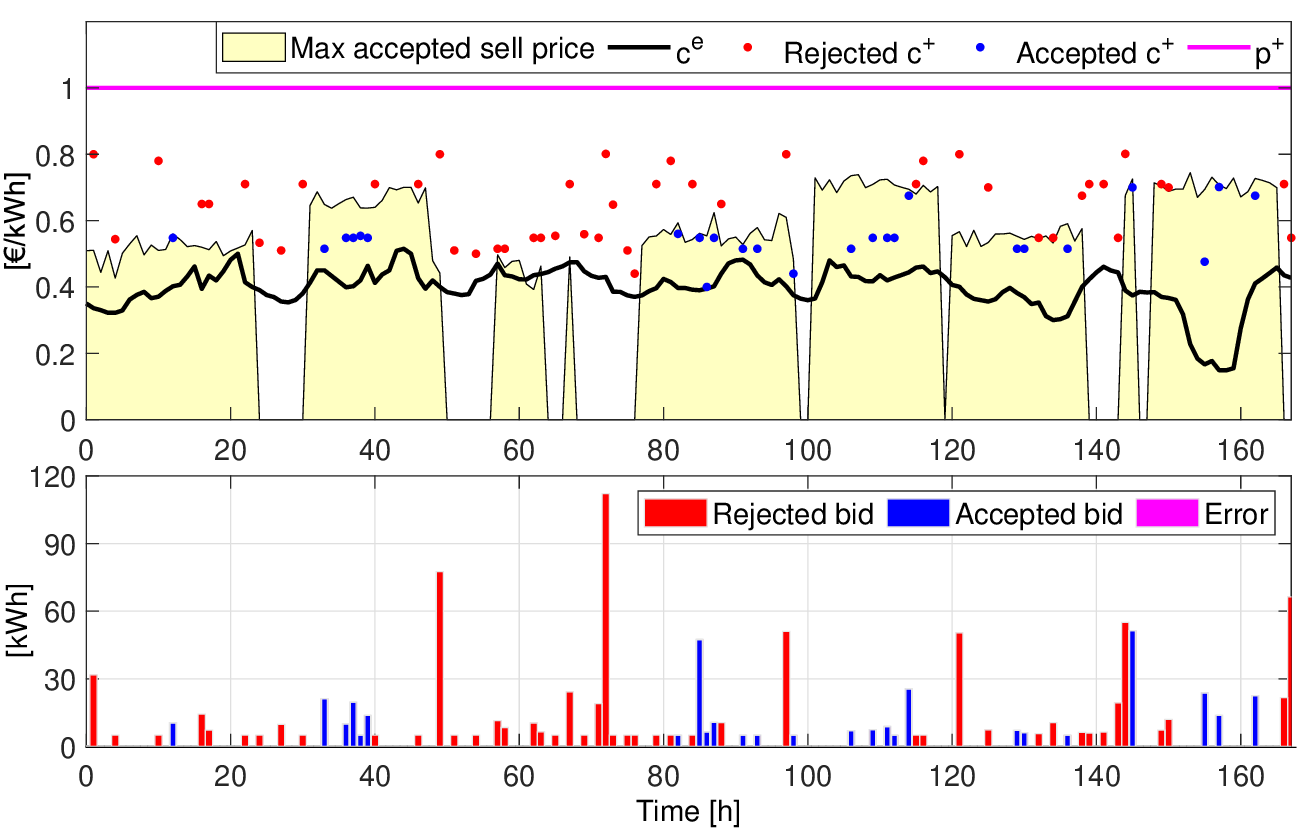}\vspace{-5pt}
    	\caption{Results: \ac{rec} sell bids: prices (top) and quantities (bottom).}
	\label{fig:sell_bids}
\end{figure}

\begin{figure}[t]
	\centering
    \includegraphics[width=0.9\columnwidth]{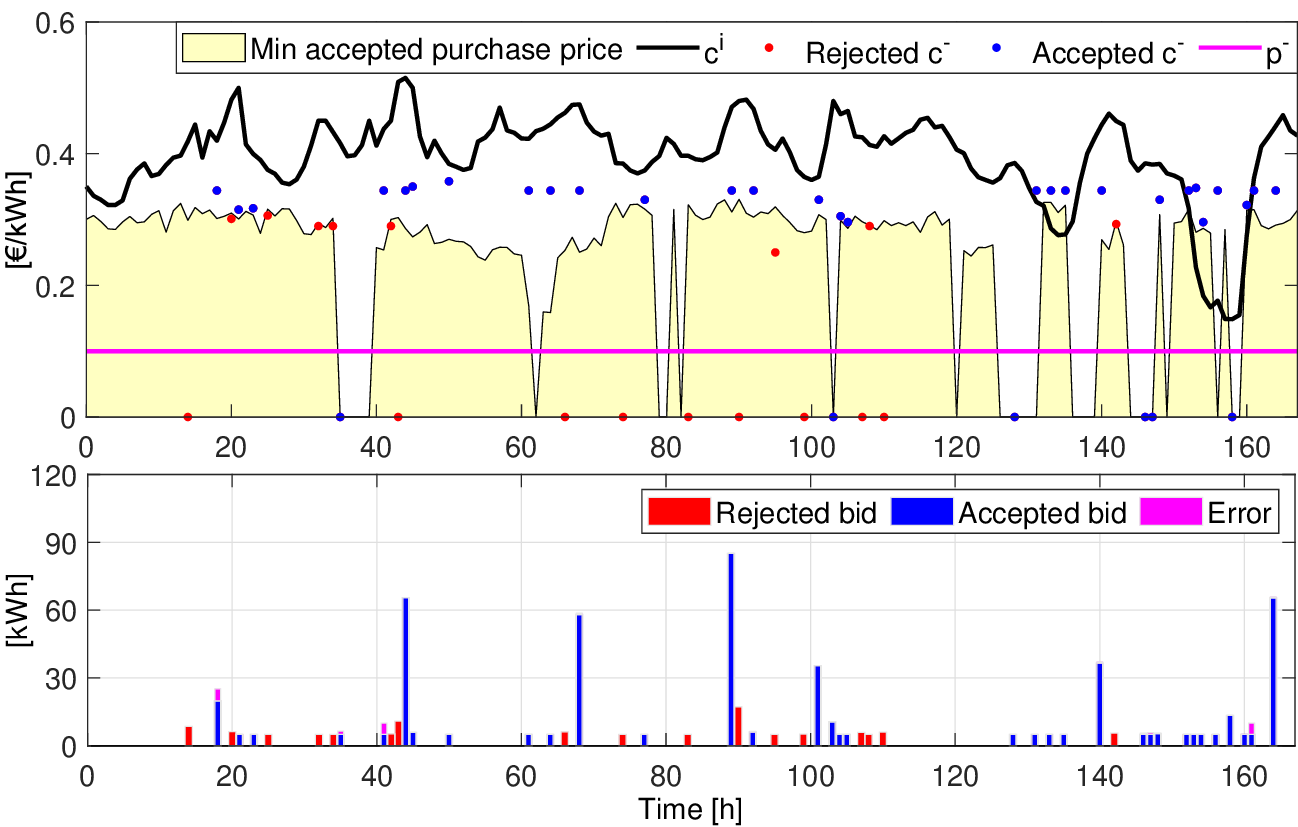}\vspace{-5pt}
    	\caption{Results: \ac{rec} purchase bids: prices (top) and quantities (bottom).}
	\label{fig:purchase_bids}
\end{figure}

The results of \ac{msd} participation are depicted in Fig.~\ref{fig:sell_bids} and Fig.~\ref{fig:purchase_bids}. Fig.~\ref{fig:sell_bids} illustrates the sell bids presented by the \ac{rec} during the simulated week. When bids are accepted, the \ac{rec} is able to provide the service with no error $E^+_k$. In general, we observe that sell bid prices are often significantly higher than $c^e_k$. This means that the optimization considers an extra discharge of the \ac{bess} cost-effective only at prices significantly higher than $c^e_k$,  even if, in this way, the percentage of accepted bids is lower.    
In Fig.~\ref{fig:purchase_bids} we observe that purchase bids are often higher than $\underline{c}^-_{k}$, leading to a high percentage of acceptance. This means that the optimization frequently considers extra charge of the \ac{bess} cost-effective. 
Few times a non zero error $E^-_k$ occurred mainly due to the errors in the prediction of \ac{pv} generation, that affects the available charging power according to \eqref{eq:green}.

\begin{figure}[t]
	\centering
    \includegraphics[width=0.9\columnwidth]{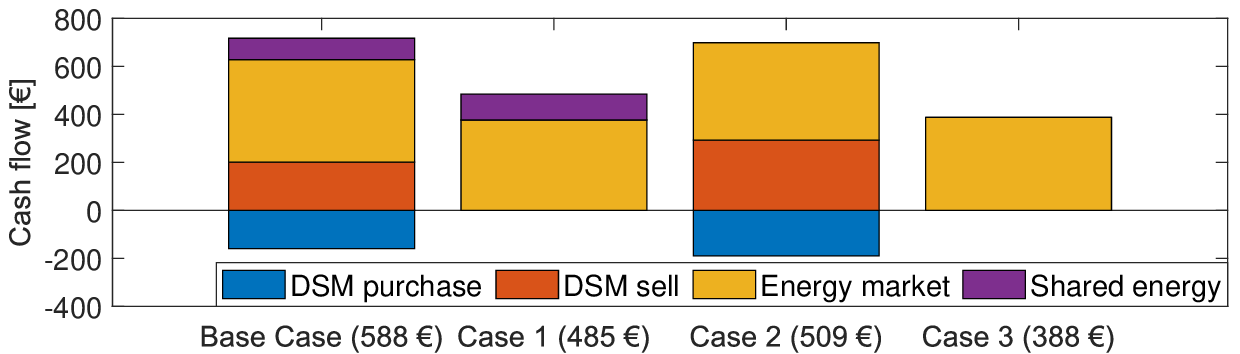}\vspace{-5pt}
    	\caption{Results: \ac{rec} cash flows. In brackets the total cash flow.}
	\label{fig:cash_flows}
\end{figure}

The one week simulation has been repeated in three different scenarios: Case~1, where the \ac{rec} does not participate to the \ac{msd}; Case~2, where the incentive for shared energy $\gamma^{sh}$ is neglected; Case~3, where the incentive for shared energy $\gamma^{sh}$ is neglected and the \ac{rec} does not participate in the \ac{msd}. The results are compared in terms of cash flows in Fig.~\ref{fig:cash_flows}. By this comparison it is possible to understand which is the actual advantage provided by the remuneration of shared energy and by the participation in \ac{msd}. Starting from Case~3: in Case~1 we obtain a cash flow increase of 25\%, due to the sharing energy incentive; in Case~2 we get a cash flow increase of 31.2\%, proving the profitability of the \ac{msd} participation. As expected the largest cash flow is realized in the base case, 21.2\% more than Case~1, and 15.5\% more than Case~3.

\section{Conclusions}\label{sec:Conclusions}
In this paper we have proposed a scenario-based day-ahead optimization algorithm for a \ac{rec} participating in the Italian pay-as-bid ancillary service market. The algorithm is able to state the \ac{rec} energy exchange baseline and decide the sell and purchase bids to be submitted to the service market with the objective of maximizing the \ac{rec} cash flow. Simulation results prove the effectiveness of the approach and the advantages for the \ac{rec} in participating in the service market.

\vspace{.2cm}

\end{document}